\documentclass[twocolumn]{aastex61}

\shorttitle{negative flare}
\shortauthors{Kobanov et al.}
\begin{document}
\title{Negative flare in the He\,\textsc{i} 10830\,\AA\ line in facula}
\correspondingauthor{N.~\surname{Kobanov}}
\email{kobanov@iszf.irk.ru}
\author{Nikolai~\surname{Kobanov}}
\affil{Institute of Solar-Terrestrial Physics
                     of Siberian Branch of Russian Academy of Sciences, Irkutsk, Russia}
\author{Andrei~\surname{Chelpanov}}
\affiliation{Institute of Solar-Terrestrial Physics
                     of Siberian Branch of Russian Academy of Sciences, Irkutsk, Russia}
\author{Vasiliy~\surname{Pulyaev}}
\affiliation{Institute of Solar-Terrestrial Physics
                     of Siberian Branch of Russian Academy of Sciences, Irkutsk, Russia}
\begin{abstract}

A small-scale flare SOL2012-09-21T02:19 (B2) occurred in a spotless active region that we observed at a ground-based telescope equipped with a spectrograph. During the flare, we registered an increase in absorption in the He\,\textsc{i} 10830\,\AA\ line by 25\%, while other chromospheric and coronal spectral lines demonstrated increase in brightness at the same location. This phenomenon called negative flare had rarely been observed at the Sun before. In this paper, we describe the morphology of this flare and investigate its dynamics based on our spectral observations and space imaging data.
The H$\alpha$ and He\,\textsc{i} 10830\,\AA\ lines reach their extreme intensities 5 and 6 minutes after the 171\,\AA\ line.
The brightening first occurred in the 171\,\AA\ and 193\,\AA\ Solar Dynamics Observatory (SDO) channels followed by the 94\,\AA, 304\,\AA, and 1600\,\AA\ signals $\sim$2 minutes after (for the maximum phases). However, the abrupt changes in line-of-sight (LOS) velocities in the chromospheric lines occur simultaneously with the intensity changes in the 304\,\AA\ and 1600\,\AA\ lines: we observed a downward motion that was followed by two upward motions. The measured horizontal speed of the perturbation propagation was close to 70\,km\,s$^{-1}$ both in the chromospheric and coronal lines.

We assume that we observed the photoionization-recombination process caused by UV radiation from the transition region during the coronal flare. With this, we point out the difficulties in interpreting the time lag between the emission maximum in the SDO UV channels and the second absorption maximum in the He\,\textsc{i} 10830\,\AA\ line.

\end{abstract}
%\keywords{Sun: activity --- 
%Sun: faculae, plages --- Sun: flares}

\section{Introduction} \label{sec:intro}

Solar flare is the most dynamic energy phenomenon observed in the solar atmosphere. It has always been a topic attracting attention of researchers. Most of the early researches are based on H$\alpha$ line observations. Then the radio-range observations complemented optical observations. The beginning of the space exploration made it possible to observe the Sun in the UV and X-ray \citep{Masuda, Krucker} emission inaccessible from the ground, which further advanced solar flare studies. Later, the flare studies became multi-wave \citep{Fletcher,Hudson,Fleishman}. In the last three decades, the IR spectral range observations have advanced as well due to the instrument development. Flare observations in the He\,\textsc{i} 10830\,\AA\ line are of interest, since they allow us to complement the UV and X-ray in analyzing non-thermal effects in the flare processes \citep{2005A&A...432..699D,2006SoPh..235..107L,2007SoPh..241..301L,2008ChJAA...8..723D}.

The He\,\textsc{i} 10830\,\AA\ line is formed by the \textit{2s$^{3}$s--2p$^{3}$p} transition, and it is observed in the upper chromosphere and transition region. The structure details in the He\,\textsc{i} images look similar to those in the H$\alpha$ and Ca\,\textsc{ii}\,H images. But, unlike these chromospheric lines, the He\,\textsc{i} line reveals the structure in the coronal holes too. The chromosphere-coronal dualism of this line's properties was analyzed by \citet{Andretta}. \citet{1976SoPh...49..315L} showed, that the opacity in the He\,\textsc{i} 10830\,\AA\ line depends on the electron density. During flares, this line can show an increased emission or an increased absorption; different combinations are possible, depending on the flare power. For example, \citet{Rust} stated that most small flares produce the emission concentrated in separate kernels in this line images. Aside from the local conditions, external factors like the coronal UV radiation or energetic particle beams may influence the formation of this line. These unique properties of the He\,\textsc{i} 10830\,\AA\ line allow to diagnose the processes of the energy transfer in the lower solar atmosphere with the use of this line.

 Note that as back as 1984 \citet{1984SoPh...91..127H} described an astonishing phenomenon---darkening in the flare ribbons in the He\,\textsc{i} 10830\,\AA\ line, which they related to the X-ray flux generated by a powerful flare. In their papers, \citet{1976SoPh...49..315L,1994IAUS..154...35A,1998ARep...42..819S,2014ApJ...793...87Z,2005A&A...432..699D, Centeno, Allred, 2016A&A...594A.104L} discussed the mechanisms responsible for the increase in the He\,\textsc{i} 10830\,\AA\ line absorption and emission.

A negative flare---or a dark light flare---is a rare event in solar observations. For the last thirty years, only a few observational facts have been described (see, e.g., \citet{2016ApJ...819...89X}). \citet{2013ApJ...774...60L} observed dark flare ribbons in D3, while these ribbons were bright in the H$\alpha$ and He\,\textsc{i} 10830\,\AA\ lines. The dark ribbon had a 5\% negative contrast, and the flare perturbation moved along the polarity inversion line. In their paper, \citet{2016ApJ...833..250W} observed an increased absorption in the He\,\textsc{i} 10830\,\AA\ line in a region close to the flare location. Besides, they observed running penumbral waves and umbral flashes in absorption, amplified during the flare. The authors believe that their results directly prove the effect of the photoionization-recombination mechanism (PRM). As an additional argument, the authors stated that the enhanced absorption in the He\,\textsc{i} 10830\,\AA\ line coincided with two microflares, whose impulse emission in the He\,\textsc{i} 10830\,\AA\ and He\,\textsc{ii} 304\,\AA\ lines located at the the main flare site in the area adjacent to the sunspot. The authors of the paper \citet{2016ApJ...819...89X} observed two negative flares in the He\,\textsc{i} 10830\,\AA\ line. One of the events was a two-ribbon flare, whose one ribbon was observed in the emission, while the other one showed a negative contrast up to 13\% and moved at a speed of 3.7\,km\,s$^{-1}$. In the other flare, the front fringe in one of the ribbons was dark, and the back part was observed in the emission. The authors think that the collisional ionization-recombination scenario explains the narrow negative flare fronts better. According to the paper, the strong emission in the H$\alpha$ red wing during the initial phase of the flare and the narrow dark ribbon in the He\,\textsc{i} image in the same location unambiguously point at a short-lived effect of a beam of energetic electrons.

%Our case differs from the above examples, since the He\,\textsc{i} negative flare matched the location of the H$\alpha$, 304\,\AA, 171\,\AA, 193\,\AA, 94\,\AA\ lines brightening, and the emission in the He\,\textsc{i} line did not follow the negative flare in this line. This is probably a result of the fact that the flare is low-power and located in a spotless facular region.

Dimmings in radiowaves have been studied since the 1950-s \citep{Covington53,Covington,Kuzmenko,Grechnev}. The term `negative flare' was also used to describe dimming in microwave radiation \citep{Sawyer,Maksimov}. In this case, the decrease in the observed intensity results from an eruptive prominence that blocked the radiation. Dimmings also can be observed in X-rays. A region becomes dark after a flux rope erupts, leaving density-depleted regions behind \citep{Zarro, Krista}.

In this paper we aim to obtain composite multi-wavelength information on the the negative flare that we observe. At the same time, we focus our main attention on the observational characteristics of the upper chromosphere processes, accompanying this event.

\section{Observational Data and Methods}

The spectral observations were carried out at the Horizontal Solar Telescope in the Sayan Solar Observatory \citep{2013SoPh..284..379K,2015SoPh..290..363K}. The telescope is located at an altitude of 2000\,m above sea level; it provides spatial resolution of 1.0--1.5\,arcsec on average. The spectral resolution is 25\,m\AA\ pixel $^{-1}$ for  the He\,\textsc{i} line and 14\,m\AA\ pixel $^{-1}$ for the H$\alpha$ line. The slit of the spectrograph covers a 1.5$\times$65$''$ region on the solar disk. One pixel of the CCD camera corresponds to a 0.2 arcsec distance on the disk.

\begin{figure}[ht!]
\plotone{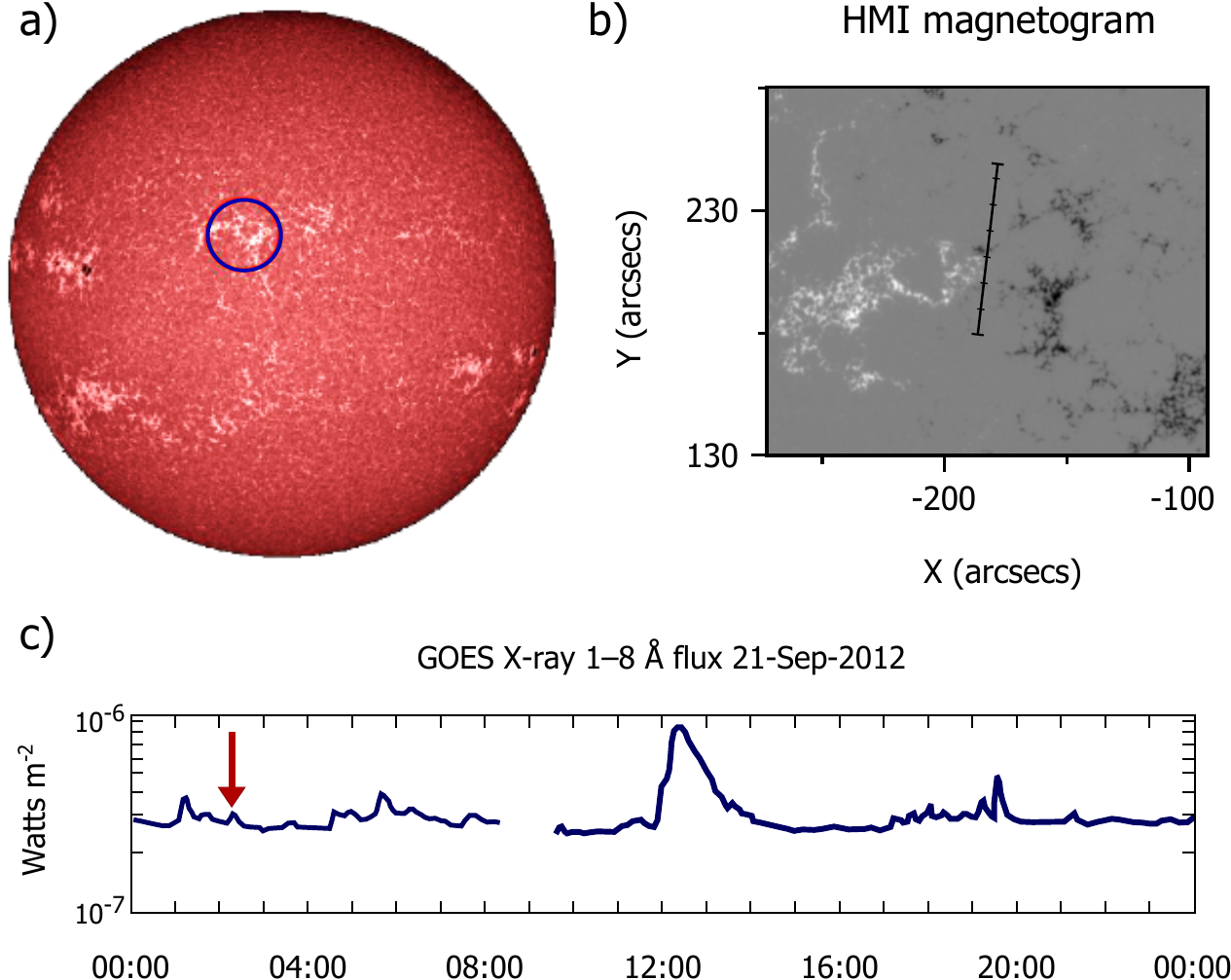}
\caption{a) the full disk in the 1700 A SDO channel on the observation date. In this band pass, faculae are seen as bright areas. The blue circle shows the observed facula. b) LOS magnetic field in the facular region. The black bar in the center shows the spectrograph slit position. c) GOES X-ray flux in the 1--8\,\AA\ range. The red arrow marks the hump associated with the flare that we observed.\label{fig:1}}
\end{figure}

The flare SOL2012-09-21T02:19 (B2) was recorded under good seeing conditions. The spotless active region 11573 that we observed consisted of a compact facular region located at 19N13E. The series time length is 100\,min; the time cadence is 1.5\,s. Along with the He\,\textsc{i} 10830\,\AA\ line, two other lines---the photospheric Si\,\textsc{i} 10827\,\AA\ line and H$\alpha$ 6563 \AA\ line---were recorded in the spectrograms.

Half-width of the spectral lines was determined as the profile width at the half intensity level between the line core and the closest continuum. Velocity signals were obtained using lambdameter technique. This method implies positioning two virtual slits with a fixed distance between them in the line wings so that the intensities in these slits are equal. The sensitivity of the method is limited only by the capability of the detector elements to distinguish small changes in the intensity caused by the line shifts, similarly to the Babcock magnetograph with wide photometer slits in the line wings. Initially, the  distance between the slits is set to correspond the wing positions of the half-intensity level in the first frame of the series averaged over the slit. In each succeeding frame, a new equal-intensity position of the slits is determined, while the distance between them remains constant. Before applying this procedure, we interpolated the line contours. This allowed us to resolve velocities down to 20\,m\,s$^{-1}$.

For the analysis, we also used data from Geostationary Operational Environmental Satellite (GOES), Reuven Ramaty High Energy Solar Spectroscopic Imager (RHESSI), and the Solar Dynamics Observatory (SDO), which provides UV images with a 12\,s or 24\,s cadence.

To coalign the ground-based and SDO observations, we first roughly identified the location during the observation when we positioned the slit on the solar image. A more precise co-alignment was done when we compared the location of the flare brightening occurrence in the 304\,\AA\ and 1600\,\AA\ channels to those in the ground-based telescope data. The location of the magnetic hill served as another reference point. As a result, we can safely assume that we coaligned the ground-based and SDO observations with an accuracy of 1--1.5$''$.

\section{Results and discussion}

\subsection{The scale and other characteristics of the event}

\begin{figure*}
\plotone{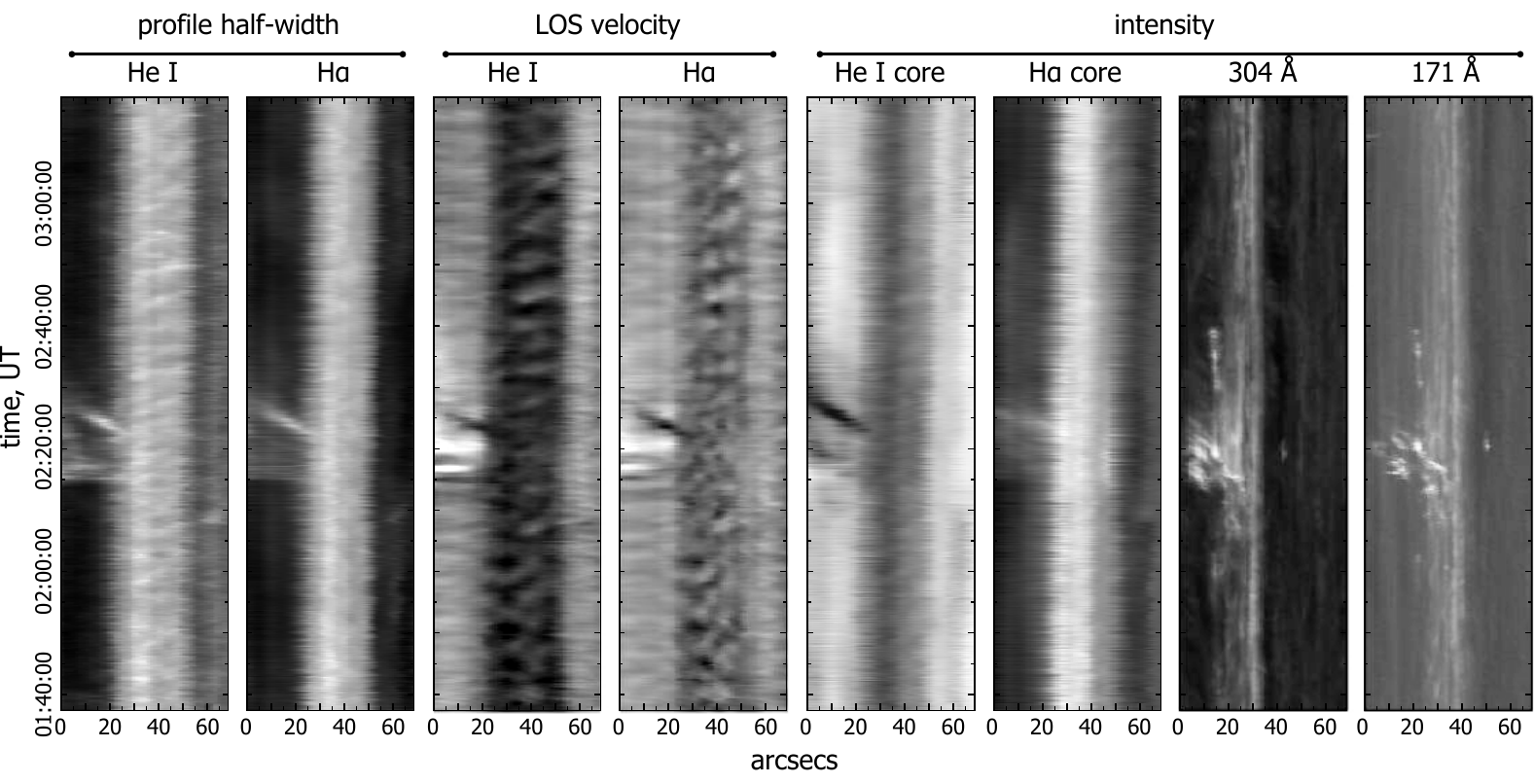}
\caption{Space-time diagrams showing the location and time of the flare. The wide vertical stripe in the center of each panel corresponds to the magnetic hill within the facular region. The flare manifests itself as the bright (or the dark in the fifth panel) stripe in the left parts of the panels. The LOS velocity diagrams show the dark and light tones for down- and upward motions.\label{fig:2}}
\end{figure*}

The faculae that we observed had a bipolar magnetic structure. Figure\,\ref{fig:1} shows the SDO magnetogram and the location of the spectrograph slit relative to the facula. The slit was located in the vicinity of the polarity inversion line, and its central part crossed a magnetic hill. The event described here located in the southern part of the slit, between the 0$''$ and 20$''$ marks (see Figure\,\ref{fig:2}).

Figure\,\ref{fig:2} shows the space-time grey-scale diagrams of line widths at the half-intensity level, line-of-sight (LOS) velocity, and intensity (core intensity for the He\,\textsc{i} and H$\alpha$ lines). The horizontal axis represents the distance along the slit in the south to north direction from 0$''$  to 68$''$. The vertical axis represents time. The distinct vertical $\sim$25$''$-wide stripe in the central parts of the diagrams results from the facular magnetic hill.

Our observations are supplemented with the diagrams, constructed from a virtual slit placed upon SDO images of the 304\,\AA\ and 171\,\AA\ channels. The  time and location of the cut are coaligned with the ground-based observations.

The flare that we refer to can be seen in the left parts of these diagrams, in the middle of the vertical axis. It manifests itself as a drastic change in all the parameters of the chromospheric lines (their intensity, LOS velocity, and line profile width) in the southern part of the slit at around 02:15. The flare was observed during 10--15 minutes, depending on the spectral line and position along the slit.

To estimate the scale of the event, we refer to the GOES observations. This was a minor flare, and it has not been registered in any flare catalog. A small hump corresponds to this flare in the GOES X-ray plot (Figure\,\ref{fig:1}). Ten hours later, a B\,8.7-class flare occurred in this facula.

A curious detail caught our attention: while the flare expressed itself as an increase in brightness in all the lines (which is always typical of flares), it appears darkened in the He\,\textsc{i} 10830\,\AA\ line (see Figures\,\ref{fig:2}, \ref{fig:5}). We ruled out the possibility of dark filament material moving into the FOV, since the location, shape, and the duration of the flare in the He I line closely repeats those in the other lines.

This darkening, or a negative flare, is a unique phenomenon. Few such cases have been observed and described before (see Section\,\ref{sec:intro}). As a rule, they were observed during powerful and medium-strength flares in the active regions with sunspots. Usually, an increase in brightness follows a negative flare, which we did not observe in this case.

\begin{figure*}
\plotone{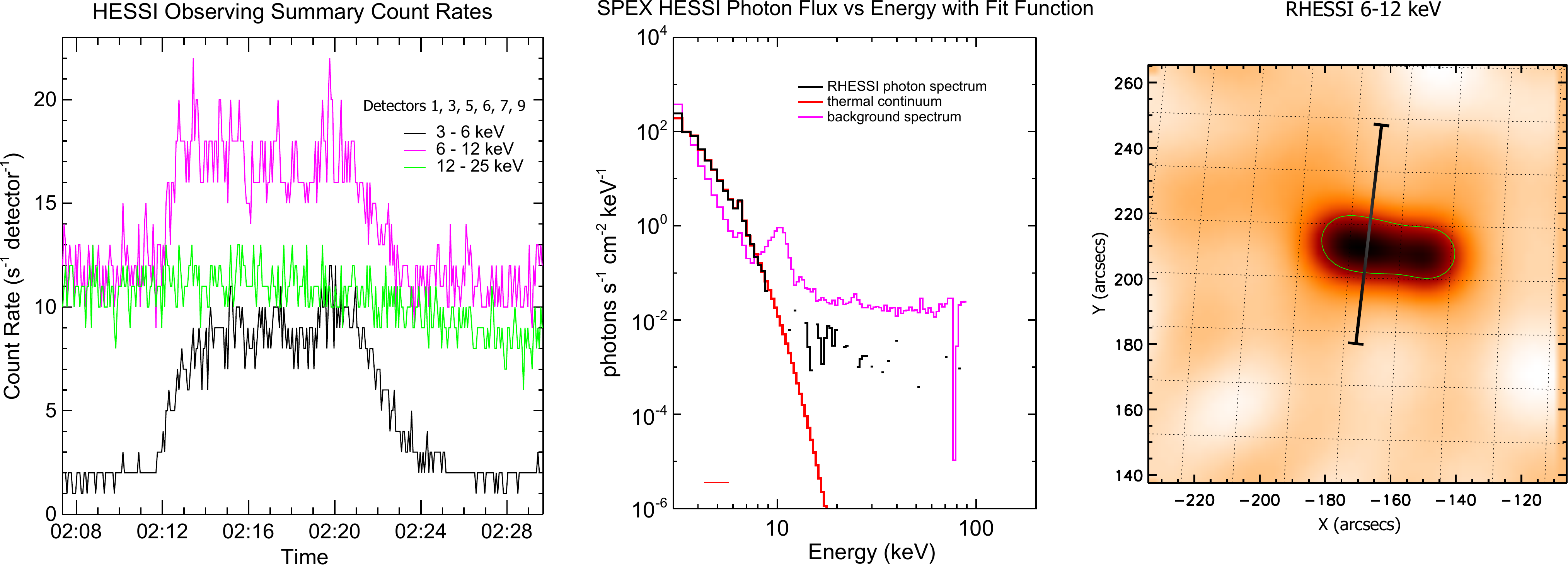}
\caption{\textit{Left:} The light curves of X-ray emission with a 4-second temporal resolution. \textit{Center:} The X-ray photon spectrum integrated between 02:12:30 and 02:21:50 UT. The black line is the observed spectrum, the magenta line is background, and the red line shows the thermal radiation function. \textit{Right:} the 6--12\ keV image of the flare region. The contour shows the half-maximum intensity level.\label{fig:6}}
\end{figure*}

We estimated the energy of the event, based on the RHESSI data \citep{rhessi1, rhessi2}. The fitting results of the photon spectrum show that X-ray emission has the thermal component only (Figure \ref{fig:6}). The fitting by the one-component thermal bremsstrahlung radiation function yields the electron temperature $T_{e}$ $=9.7\,MK$ and emission measure of $5.4\times$ 10$^{45}$\,cm$^{-3}$\. 
Thus, we had to estimate the thermal energy only. We used the same formula that \citet{Emslie} did. The area of the flare source (\textit{S}) is derived from the 6--12\,keV image at the level of 50\,\% of the maximum intensity. The volume (\textit{V}) is calculated as $S^{3/2}$, which gives $\textit{V}=5.8\times$ 10$^{27}$\,cm$^{-3}$. We assumed that the projection factor is 1.

Under these assumptions, the resulting energy of the flare is $2.2\times$ 10$^{28}$\,erg.

\subsection{Variations of the core intensity, line half-width, and equivalent width in the He\,\textsc{i} and H$\alpha$ lines}

\begin{figure}
\plotone{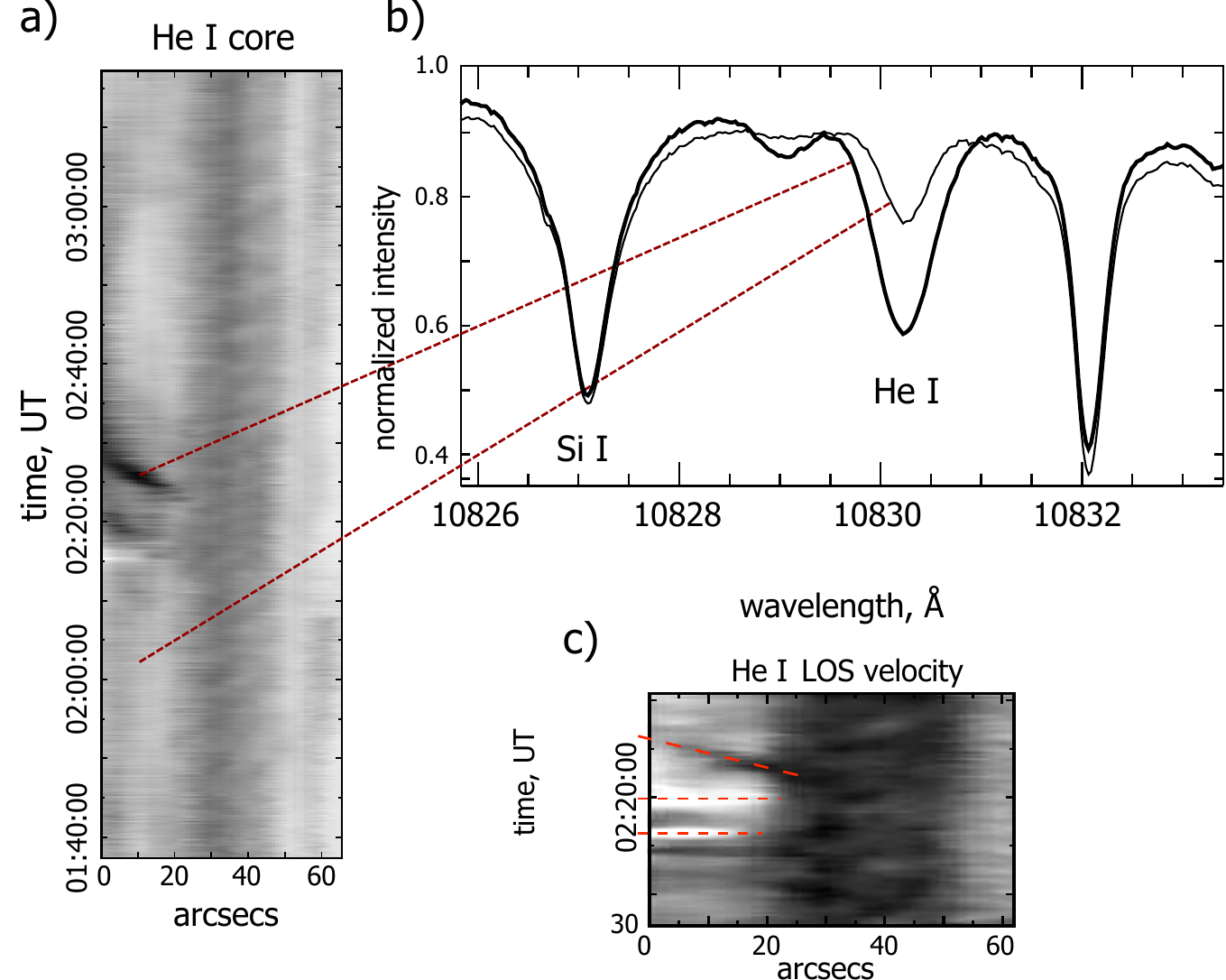}
\caption{\textit{a}: space-time diagram of the He\,\textsc{i} 10830\,\AA\ line core intensity. \textit{b}: spectra in the He\,\textsc{i} line vicinity before and during the flare. \textit{c}: He\,\textsc{i} LOS velocity space-time diagram. The dashed lines show the inclination of the stripes.\label{fig:4}}
\end{figure}

During the flare, the He\,\textsc{i} line core intensity showed two dips at 02:19 and 02:26 (Figure\,\ref{fig:4}). The H$\alpha$ line core intensity grew from 02:15 and plateaued at 02:19, then at 02:25 reached its maximum. The He\,\textsc{i} line core intensity in its minimum dropped by 25\% compared to the pre-flare state (Figure\,\ref{fig:4}), while the H$\alpha$ line core intensity increased by 8\%, and the 304\,\AA\ channel intensity increased more than 10-fold. The H$\alpha$ line half-width resembles the H$\alpha$ light curve with the maximum at 02:26. The He\,\textsc{i} profile half-width signal shows two maxima: at 02:19 and 02:26. The equivalent width of the He\,\textsc{i} line in this spatial position changes from 85\,m\AA\ to 240\,m\AA\ simultaneously with the half-width curve (Figure\,\ref{fig:5}), while the equivalent width of this line in the facular magnetic hill (33$''$\,position at the slit) is 170\,m\AA. Aside from the increase in the line's main component depth, the flare spectrograms reveal a deepened blue component profile at 10829\,\AA, which is almost flat in this region before and after the flare (Figure\,\ref{fig:4})

\subsection{Velocities in the image plane}

In the diagrams of Figure\,\ref{fig:2}, the flare in its maximum phase looks like a stripe set at an angle to the horizontal axis. Measuring this angle allows us to calculate the horizontal speed of the perturbation propagation. However, the white stripes at around 02:20 in the velocity panels look horizontal (Figure\,\ref{fig:4}c), which, under this approach, would result in an infinite propagation speed. We believe that these stripes reflect the situation when either a flare loop broadens, or the matter moves up along the line of sight simultaneously at a 20$''$-length segment.

The inclination of the stripes in the He\,\textsc{i} and H$\alpha$ panels yields a speed of 65$\pm$8 \,km\,s$^{-1}$. The SDO 304\,\AA\ and 171\,\AA\ images analysis gives a similar brightening propagation speed of 70$\pm$13 \,km\,s$^{-1}$ calculated along the direction of the perturbation propagation rather than along the slit.

\subsection{Vertical propagation lags}

\begin{figure}
\plotone{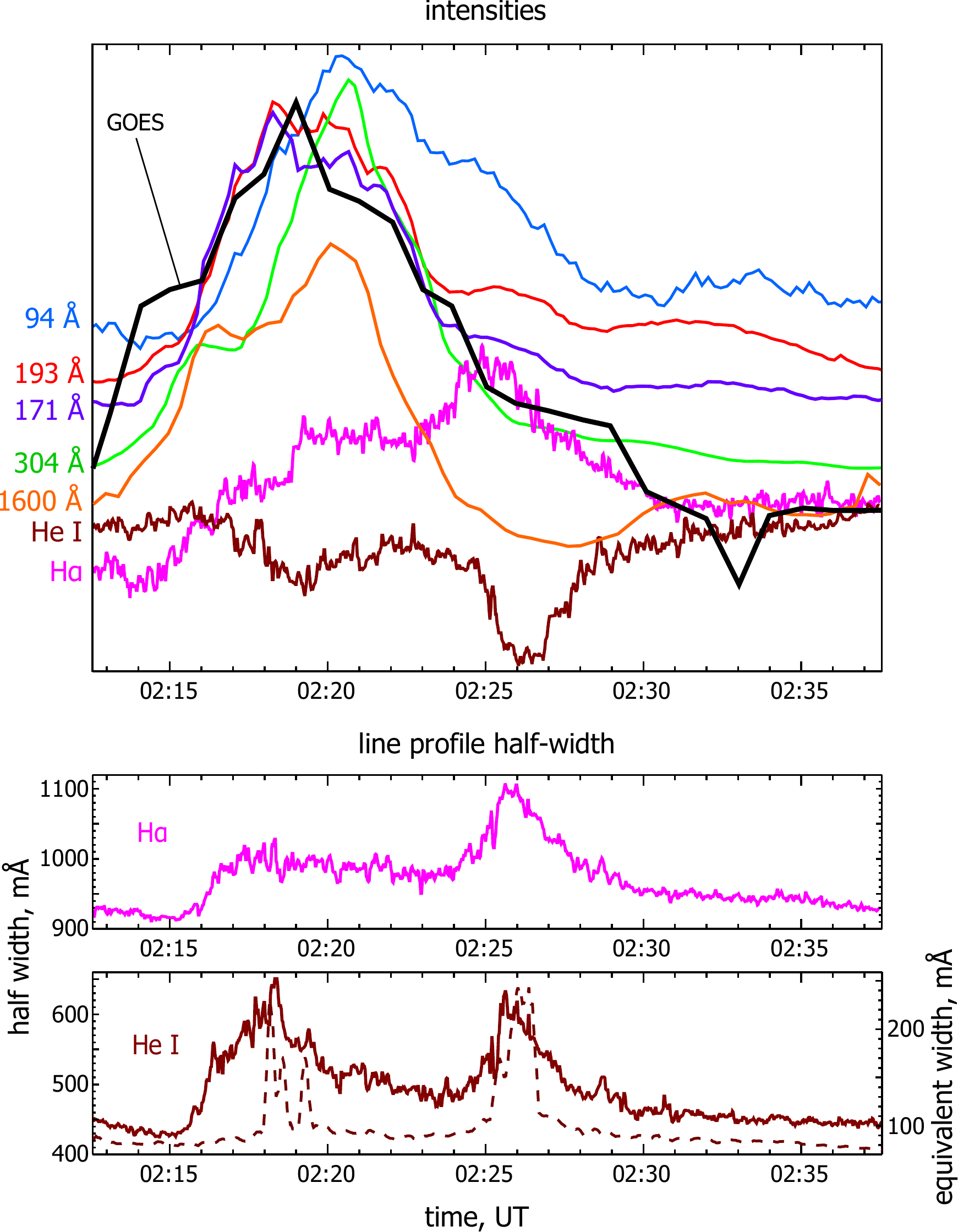}
\caption{Light curves for all the lines during the flare. Bottom panels: profile half-width variations of the He\,\textsc{i} and H$\alpha$ lines at the 10$''$ point. The dashed line in the lower panel shows the equivalent width of the He\,\textsc{i} line. \label{fig:5}}
\end{figure}

The increase in the RHESSI channels intensity starts at 02:12 (Figure\,\ref{fig:6}). Among the AIA channels, the flare first reached its maximum in the 171\,\AA\ and 193\,\AA\ channels  (Figure\,\ref{fig:5}). From these heights, the flare perturbation propagated upwards and downwards. One hundred and forty seconds after these first two channels, the 94\,\AA\ signal (associated with a greater height) peaked. It was shortly after followed by the 304\,\AA\ and 1600\,\AA\ signals, formed in the transition region and in the chromosphere respectively.

In the chromospheric H$\alpha$ and He\,\textsc{i} lines, the change in brightness gradually started developing together with the 304\,\AA\ channel, and the main peaks (brightening in H$\alpha$  and absorption in He\,\textsc{i}) lagged behind the 304\,\AA\ channel by 260\,s and 320\,s, respectively. And a sharp increase in the LOS velocity signals in these lines coincided with the flare peaks in the 304\,\AA\ and 1600\,\AA\ channels (Figure\,\ref{fig:5}, \ref{fig:7}).

\subsection{Line-of-sight velocities}

LOS velocity space-time diagrams in the He\,\textsc{i} and H$\alpha$ lines in Figure\,\ref{fig:2} show down- (dark tone) and upward (light tone) movements in the spectrograph slit field of view. The dark wide vertical stripe in the center shows the prevailing sinking in the 25$''$--50$''$ region accompanied by 5-minute oscillations. This segment corresponds to the magnetic hill, where the field strength reaches 680\,G according to Helioseismic and Magnetic Imager (HMI) observations (Figure\,\ref{fig:1}). The dark horizontal stripe (at 02:15) in both lines denotes brief downward motion followed by two abrupt upward motions (light stripes) at around 02:18 and 02:21. The LOS velocity in the main phase of the negative flare is presented as an inclined dark stripe right after the second upward motion. Its inclination to the horizontal axis (prominent in the H$\alpha$ line as well) shows that the downward motion accompanies the perturbation propagation along the slit.

\begin{figure}
\plotone{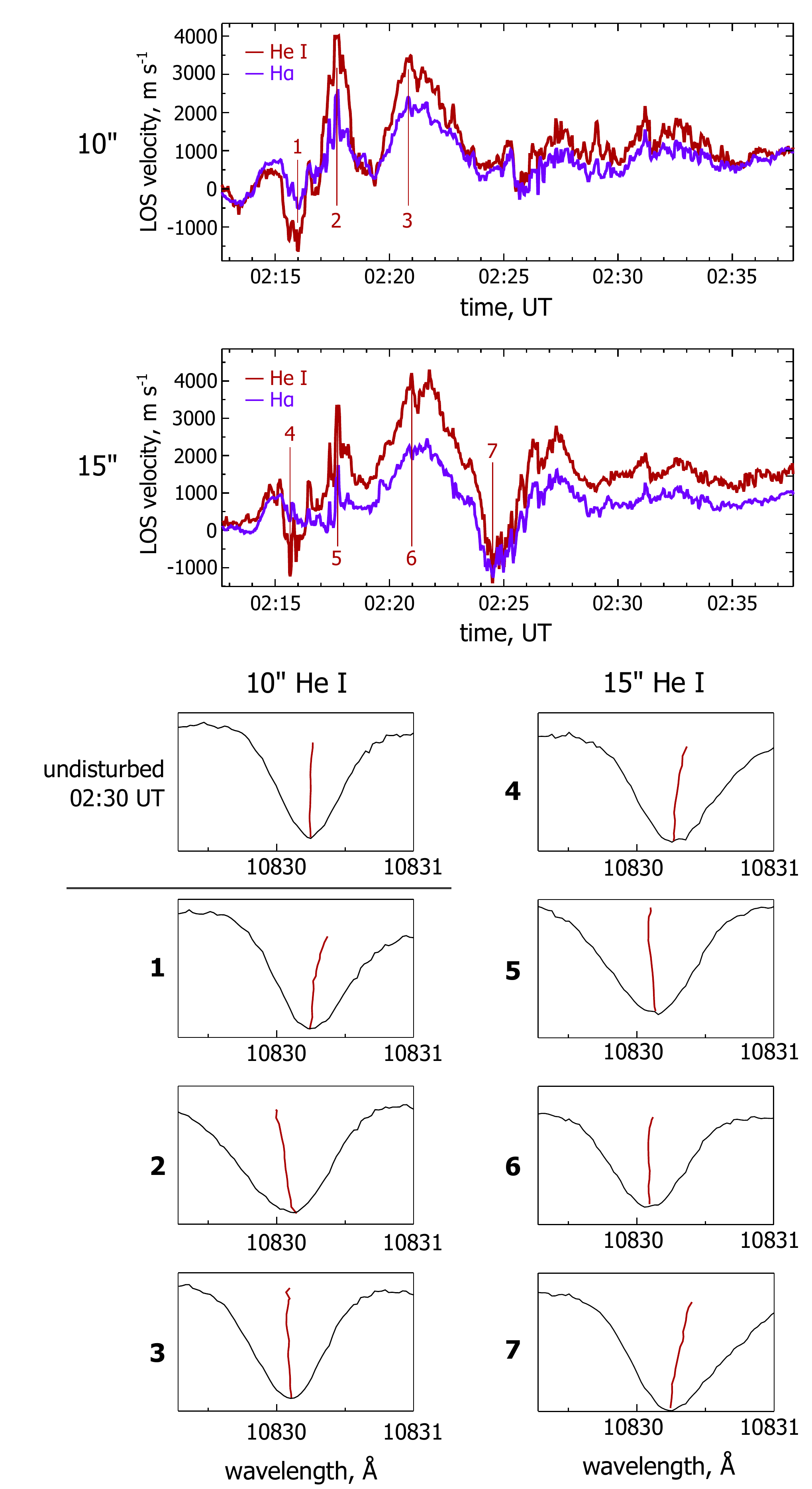}
\caption{Two upper panels: Doppler velocity signals at the 10$''$ and 15$''$ points. The LOS-velocity measurements are performed using the method described in Section 2 with a space resolution of 1.5$''$. Bottom panels: bisectors of the He\,\textsc{i} line at the 10$''$ and 15$''$ points at the extreme velocity moments marked with numerals in the upper panels. The asymmetrical profiles indicate the presence of unresolved movements.\label{fig:7}}
\end{figure}

Figure\,\ref{fig:7} shows He\,\textsc{i} LOS velocity signals at the 10$''$ and 15$''$ positions, where these signals are most pronounced. Bisector shapes in Figure\,\ref{fig:7} demonstrate that the far red wing of the He\,\textsc{i} line formed most of the downward motion signals, while the line profile shifted as a whole to the blue side during the rise. The asymmetry in the profiles may result from high LOS velocities in unresolved fine structures. The cool matter flows to the lowest layers, forming the line wings, while the hot matter moves upwards, forming the line core. The profiles in Figure\,\ref{fig:7} can be presented as a superposition of two components, one of which is the undisturbed profile observed at 02:30. At the moments marked in Figure\,\ref{fig:7}, the second blue or red component shows the shifts equivalent to velocities of over 10\,km\,s$^{-1}$. Under this approach, the motions during the flare are as follows: 02:16---downward motion at a velocity of 9\,km\,s$^{-1}$; 02:18---upward motion at 12\,km\,s$^{-1}$; 02:21---upward motion at 10\,km\,s$^{-1}$; 02:25---downward motion at 9.5\,km\,s$^{-1}$. When the matter was moving upward at 02:18 and downwards at 02:25, the half-width of the line profile increased up to 650\,m\AA\ (Figure\,\ref{fig:5}).

We cannot unambiguously interpret the complex picture of the He\,\textsc{i} profile changes in the flare location. One may assume that these changes indicate oscillations induced by the flare. Probably, an in-depth analysis will require taking into account non-equilibrium ionization effects \citep{Carlsson,Golding}, which is, however, beyond the scope of this paper.

\subsection{Discussion on the probable scenarios}

A visual analysis of the half-tone diagrams in Figure\,\ref{fig:2} gives the impression that we observe a surge from the 33$''$ point, where the facular magnetic hill is located. This impression is supported by the fact that the intensity, LOS velocity, and line profile width in the short-term surge are close to the average values at the 33$''$ point in the center of the facula (Figure\,\ref{fig:2}). At the 33$''$ point, however, these values undergo no disturbance, neither during nor before the flare. We revealed no changes in the photospheric magnetic field. Besides, in the supposed `surge', the absorption and half-width are higher than those at the 33$''$ point (by 16\% and 22\%, respectively). At the same time, we registered a pronounced downward motion, which is not typical of surges. A visual analysis of the 1600\,\AA\ and 304\,\AA\ images (the closest to the chromospheric layers) revealed no surge-like structures.

On the other hand, two impulse upward motions preceding the negative flare may indicate the chromospheric evaporation scenario \citep{1987ApJ...317..956S,2011A&A...526A...1D,2017ApJ...841L...9L}. Let's consider the possibility of this scenario. The main observational signature of chromospheric evaporation is an upward hot plasma flow during the impulse flare phase. The velocities measured in hot coronal lines reach hundreds of km\,s$^{-1}$ \citep{2013ApJ...762..133B}. At the same time, a downward cool matter flow of a much lower velocity is observed in the transition region and upper chromosphere in some cases (in powerful flares). Particle beams and thermal conduction are usually suggested as possible reasons for chromospheric evaporation. Chromospheric evaporation may be explosive or gentle. The former is observed during the impulse phases of powerful flares and is usually related to accelerated particle beams, while gentle evaporation is observed during gradual phases. At the same time, red-shifted matter in a loop footpoint is considered to be a sign of explosive evaporation \citep{Canfield}. \citet{1987ApJ...317..956S} observed upward LOS velocity of about 10\,km\,s$^{-1}$ in H$\alpha$ during the gradual phase, which they identified as gentle evaporation. Based on the Fe\,\textsc{xv} line EIS measurements, \citet{milligan} obtained velocities of 14\,km\,s$^{-1}$ for up- and downward flows observed simultaneously in different parts of the active region during a B2 flare. He concluded that the event that he studied was an explosive evaporation caused by thermal conduction.

The succession of the matter motions in our event is not typical of chromospheric evaporation. To confidently determine this, we lack spectral observations in hot coronal lines. Unfortunately, these data are absent in the EIS (Hinode) and CDS (SOHO) archives.

Recently, methods have been developed that allow us to distinguish between the mechanisms of the energy transfer to the lower atmosphere in flares. The scenario proposed by \citet{2005A&A...432..699D} suggests that during the flare's impulse phase, non-thermal electrons produce a significant increase in the He\,\textsc{i} 10830\,\AA\ line absorption, and after that they provide an increase in the emission in this line in the flare maximum phase. These variations should be much lower without non-thermal factors. According to the authors of this scenario, strong absorption in the impulse phase and strong emission in the main phase unambiguously indicate a non-thermal source of the perturbation. In the event that we observed, the absorption increased moderately at 02:18 (impulse phase?) and increased significantly at 02.26 (gradual phase). The increase in the half-width of the He\,\textsc{i} line profile from 450\,m\AA\ to 650\,m\AA\  accompanied both these moments. We observed no sign of emission in the He\,\textsc{i} line during the flare. Thus, the above scenario  is hardly applicable to our case.

\citet{Brosius} proposed to use the temporal properties of the AIA\,EUV light curves in order to identify the dominant mechanisms for energy transfer to the chromosphere during a solar flare. According to their hypothesis, if the 94\,\AA\ channel intensity increases before the 171\,\AA\ channel intensity, the thermal conduction dominates. They state that simultaneous increase in emission of all the AIA\,EUV channels implies the presence of non-thermal particles.

In our case, the temporal relationships between the AIA light curves drastically differ from those in \citet{Brosius} for the microflare B\,4.8 (July 31, 2010, 05.00--05.45). As opposed to their case, the 94\,\AA\ signal peaks more than two minutes after the 171\,\AA\ channel. Hence, according to \citet{Brosius}, we should rule out the thermal conduction as the dominant mechanism. On the other hand, all the AIA curves in Figure\,\ref{fig:5} show almost simultaneous increase in emission during the impulse phase of the flare (at 02:16). This, as their hypothesis suggests, indicates the presence of the high-energy particle beams. However, the analysis of the energy spectrum in Figure\,\ref{fig:6} based on RHESSI data showed the absence of energetic particle beams in the event. Thus, an attempt to explain our event using the criteria proposed by \citet{Brosius} leads to a contradiction.

\citet{Andretta} suggested a method to identify the agent responsible for the He\,\textsc{i} 10830\,\AA\ absorption increase: an anti-correlation between the He\,\textsc{i} 584\,\AA\ and He\,\textsc{i} 10830\,\AA\ images unambiguously indicates that photoionization-recombination mechanism forms both lines, while a positive correlation shows that the collisional mechanism (CM) dominates. Unfortunately, the absence of the He\,\textsc{i} 584\,\AA\ observational data for this event prevented us from using this method.

Note that the layer closest to the chromosphere is the transition region represented by the He\,\textsc{ii} 304\,\AA\ line. \citet{2016A&A...594A.104L} noted the special role of this region in the He\,\textsc{i} 10830\,\AA\ line absorption. In our case, during the flare, the brightness in the 304\,\AA\ channel increased by more than 10 times compared to the background level. The first darkening (02:18--02:21) in the He\,\textsc{i} light curve coincides with the intensity maximum in the He\,\textsc{ii} 304\,\AA\ channel in Figure\,\ref{fig:5}. Note that the He\,\textsc{ii} 304\,\AA\ line's blue wing is blended by the Si\,\textsc{xi} line. Its intensity, however, is 15 times lower than that of the He\,\textsc{ii} line \citep{Brosius2008}. This line may influence velocity measurements in the He\,\textsc{ii} 304\,\AA\ line \citep{Hudson2011}, but it hardly appreciably changes the light curve of the 304\,\AA\ channel. This synchroneity gives the impression that the UV radiation from the adjacent transition region caused the increased absorption in the He\,\textsc{i} line. We, however, fail to explain that the He\,\textsc{i} absorption reached its maximum 6 minutes after the He\,\textsc{ii} 304\,\AA\ did. We may assume that the increase in the equivalent width and absorption in the He\,\textsc{i} line core (Figure\,\ref{fig:5}) were caused by an increase in the chromosphere density \citep{1976SoPh...49..315L,2016A&A...594A.104L}, which, in turn, resulted from the cool matter downflow at 02:25 (Figure\,\ref{fig:7}). But in this case the reason for the brief downflow 6--7\,min after the maximal brightness in the corona remains unclear as well.

%In general, the temporal characteristics of the light curves in Figure\,\ref{fig:5} agree with no scenario listed in the literature.

%Note that in faculae, the He\,\textsc{i} 10830\,\AA\ line depth increases several times compared to the surrounding quiet Sun \citep{1963BAN....17...93N}. In the case of small faculae, \citet{2013SoPh..284..379K} used this effect to control the position of the observed object at the spectrograph slit. By the way, \citet{2016A&A...594A.104L} stated that `vertically oriented chromospheric structures that are surrounded by hot, dense, transition region and coronal material will typically produce the strongest line'. In our opinion, this definition describes the facular magnetic hill, whose LOS magnetic field maximum indicates a similar vertical structure. Note that \citet{2016A&A...594A.104L} did not consider their calculations representative of active regions and phenomena like faculae and flares.

The temperature approach may help establish the agent causing the absorption in this flare. The collisional mechanism requires a temperature of higher than 20\,000\,K \citep{Andretta, 2005A&A...432..699D}, while a temperature of about 10\,000\,K is sufficient for the photoionization-recombination mechanism. Using the dependence of H$\alpha$ Doppler width on the temperature, we roughly estimated the temperature in the chromosphere to be 10\,000 to 15\,000\,K under the assumption of exclusively temperature-related widening. This justifies the choice of PRM over CM as the agent causing the absorption in this flare.

In any case, however, the event that we observe is related to the small-scale flare in the corona and is caused by it.

\section{Conclusions}

The key point of this study is that we observed a sharp increase in absorption in the He\,\textsc{i} 10830\,\AA\ line by 25\% and drastic changes in the main characteristics of this line during a microflare. It was accompanied by an increase in intensity in the H$\alpha$, 1600\,\AA, 304\,\AA, 171\,\AA, 193\,\AA, and 94\,\AA\ lines. In the chromosphere, the flare manifested itself as drastic changes in the equivalent width, intensity, and LOS velocity. The equivalent width of the He\,\textsc{i} line increased from 85 to 240\,m\AA.  We identified this event as a negative flare.

The asymmetry of the He\,\textsc{i} profile changing during the flare indicates an unresolved movements. Using the two-component profile modelling, we determined the directions and velocities of these movements. Downward motion having a LOS velocity of 9\,km\,s$^{-1}$ and two upward motions having a LOS velocity of 10--12\,km\,s$^{-1}$ preceded the negative flare in the chromosphere. The He\,\textsc{i} absorption maximum coincided with the downward motion with the LOS velocity reaching 9.5\,km\,s$^{-1}$.

We measured horizontal speed of the perturbation propagation in the chromosphere, transition region, and corona. In the chromosphere, the speed along the slit was 65\,km\,s$^{-1}$, while the speed of the perturbation in the 304\,\AA\ and 171\,\AA\ channels was 70\,km\,s$^{-1}$.

The time differences between the intensity maxima at different heights are as follows: there is no time lag between the 171\,\AA\ and 193\,\AA\ lines, where the flare first occurred. One hundred and forty seconds later, the flare developed higher in the 94\,\AA\ line, while one hundred and sixty seconds later it peaked lower in the 304\,\AA\ and 1600\,\AA\ lines. The intensity increase in these lines coincided with short-term variations of the LOS velocities in the chromospheric H$\alpha$ and He\,\textsc{i} lines. The intensities, however, reached their extreme values in the H$\alpha$ and He\,\textsc{i} lines only 260\, and 320\,s later.

The ensemble of the event's characteristics does not correspond in detail to any of the scenarios listed in the literature. We assume that we observed the photoionization-recombination process caused by the UV radiation from the transition region during the coronal flare. This conclusion is supported by the lack of evidence for the heightened chromosphere temperature in the event location. With this, we point out the difficulties in interpreting the time lag between the emission maximum in the He\,\textsc{ii} 304\,\AA\ channel and the second absorption maximum in the He\,\textsc{i} 10830\,\AA\ line. Probably, to interpret such events more comprehensively, we need to complement spectral observations with He\,\textsc{i} and H$\alpha$ image series with higher spatial resolution.

\acknowledgments

The research was partially supported by the Projects No.\,16.3.2 and 16.3.3 of ISTP SB RAS, and by the Russian Foundation for Basic Research under grant No.\,16-32-00268 mol\_a. Spectral  data on the chromosphere were recorded at the Angara Multiaccess Center facilities at ISTP SB RAS. We are grateful to the NASA/SDO and RHESSI science teams for providing the data. We would like to thank L.~K.~Kashapova for the help in calculating the energy of the event. We would also like to thank V.~V.~Grechnev, V.~M.~Grigoryev, A.~V.~Borovik, and J.~Leenaarts for their helpful remarks and suggestions, as well as E.~N.~Korzhova for her help in preparation of the manuscript.

\end{document}